\documentclass[11pt]{revtex4}    
\usepackage{amssymb,epsf}
\usepackage{latexsym}
\begin{document}

\title{The relation between $F(R)$ gravity and\\
Einstein--conformally invariant Maxwell source}
\author{S. H. Hendi\footnote{hendi@mail.yu.ac.ir}}

\address{Physics Department,
College of Sciences, Yasouj University, Yasouj
75914, Iran\\
Research Institute for Astrophysics and Astronomy of Maragha
(RIAAM), P.O. Box 55134-441, Maragha, Iran}

\begin{abstract}
In this paper, we consider the special case of $F(R)$ gravity, in
which $F(R)=R^{N}$ and obtain its topological black hole solutions
in higher dimensions. We show that, the same as higher dimensional
charged black hole, these solutions may be interpreted as black
hole solutions with two event horizons, extreme black hole and
naked singularity provided the parameters of the solutions are
chosen suitably. But, the presented black hole is different from
the standard higher-dimensional Reissner-Nordstr\"{o}m solutions.
Next, we present the conformally invariant Maxwell field coupled
to Einstein gravity and discuss about its black hole solutions.
Comparing these two class of solutions, shows that there is a
correspondence between the Einstein-conformally invariant Maxwell
solutions and the solutions of $F(R)$ gravity without matter field
in arbitrary dimensions.
\end{abstract}

\maketitle

\section{Introduction}

Various proposals of diverse characters have been suggested by the
physicists during the past decades for going beyond, or modifying,
Einstein general relativity and often for few viable reasons. The
most important motivation coming from high-energy physics for
adding higher order invariants to the gravitational action, as
well as a general motivation coming from cosmology and
astrophysics \cite{Capozziello} for seeking generalizations of
Einstein gravity. The so-called modified gravities constructed by
adding correction terms to the usual Einstein-Hilbert action (e.g:
\cite{FR,Carroll,Lovelockpaper}), is only one endeavor among
others to go beyond Einstein general relativity and have opened a
new window to study the origin of the current accelerated
expansion of the Universe \cite{Expansion1,Expansion2,Expansion3}.
For example in Lovelock gravity, there have been some attempts for
understanding the role of the higher curvature terms from various
points of view \cite{Lovelockpaper}. As an another example, a
special case of considering the effect of higher curvature
corrections, we will deal with in this paper the so-called $F(R)$
gravity \cite{FR} (and for a review, see, \cite{Sotiriou}), whose
action is an arbitrary function of curvature scalar $R$. When
$F(R)=R$, the Einstein's general relativity is recovered. Some of
the main reasons to consider $F(R)$ gravity are as follows: First
of all, there is simplicity: $F(R)$ actions are sufficiently
general to encapsulate some of the basic characteristics of
higher-order gravity, but at the same time they are simple enough
to be easy to handle. Second, there are serious reasons to believe
that $F(R)$ theories are unique among higher-order gravity
theories, in the sense that they seem to be the only ones which
can avoid the long known and fatal Ostrogradski instability
\cite{Wood}. Third, $F(R)$ theories have no ghosts
\cite{Buchbinder}, and the stability condition $F^{\prime \prime
}(R)\geq 0$ of \cite{Dolgov} essentially amounts to guarantee that
the scalaron is not a ghost.

In $F(R)$ gravity, Einstein equations posses extra terms induced
from geometry which, when moved to the right hand side, may be
interpreted as a matter source represented by the energy-momentum
tensor $T_{\mu \nu}^{Curv}$, see equation (\ref{FE},\ref{Tcurv}).
In a similar fashion, the Space-Time-Matter ($STM$) theory,
discussed below, results in Einstein equations in $d-$dimension
with some extra matter terms showed by the energy-momentum tensor
$T_{\mu \nu}^{matt}$. It therefore seems plausible to make a
correspondence between the matter terms in $STM$ theory and
geometrical terms in $T_{\mu \nu}^{Curv}$ resulting in $F(R)$
gravity.

From the other point of view, straightforward generalization of
the electromagnetic field to higher dimensions one essential
property of it is lost, namely, conformal invariance. In Ref.
\cite{Zhuk}, it has been shown that the conformal excitations of
the extra-dimensional space components have the form of massive
scalar fields living in the external (our) spacetime. Maxwell
theory can be studied in a gauge which is invariant under
conformal rescalings of the metric, and firstly, has been proposed
by Eastwood and Singer \cite{EasSin}. Also, there exists a
conformally invariant extension of the Maxwell action in higher
dimensions (Generalized Maxwell Field, GMF), if one uses the
lagrangian of the $U(1)$ gauge field in the form
\cite{HasMar1,HasMar2,HasMar3,HendiRastegar,HendiPLB}
\begin{equation}
I_{GMF}=\kappa \int d^{n+1}x\sqrt{-g}\left( F_{\mu \nu }F^{\mu \nu }\right)
^{s},  \label{IGMF}
\end{equation}
where $F_{\mu \nu }=\partial _{\mu }A_{\nu }-\partial _{\nu
}A_{\mu }$ is the \ Maxwell tensor and $\kappa $ is an arbitrary
constant. It is straightforward to show that for $s=(n+1)/4$, the
action (\ref{IGMF}) is invariant under conformal transformation
($g_{\mu \nu }\longrightarrow \Omega ^{2}g_{\mu \nu }$ and $A_{\mu
}\longrightarrow A_{\mu }$) and for $n=3 $, the action
(\ref{IGMF}) reduces to the Maxwell action as it should be. The
idea is to take advantage of the conformal symmetry to construct
the analogues of the four-dimensional Reissner-Nordstr\"{o}m black
hole solutions in higher dimensions.

The main scope of this work is to present the correspondence between $F(R)$
gravity and Einstein-conformally invariant Maxwell theory. As we show later,
these solutions have some interesting properties, specially in the
electromagnetic fields, which do not occur in Einstein gravity in the
present of ordinary Maxwell field.

The outline of our paper is as follows. In section \ref{FieldF(R)}
we present a short review of field equations of
$(n+1)$-dimensional $F(R)$ gravity. In section \ref{SolF(R)} the
field equations of $F(R)=R^{N}$ gravity are solved in the absence
of matter field and the resulting solutions are interpreted as
black hole. The solutions of Einstein-conformally invariant
Maxwell source are considered in section \ref{ENonMax}.
Conclusions are drawn in the last section.

\section{ Basic Field Equations of $F(R)$ gravity:\label{FieldF(R)}}

The action of $F(R)$ gravity, in the presence of matter field has
the form of
\begin{equation}
\mathcal{I}_{G}=-\frac{1}{16\pi }\int d^{n+1}x\sqrt{-g}\left[
F(R)+\mathcal{L}_{matt}\right] ,  \label{Action}
\end{equation}
where $R$ is the scalar curvature and $F(R)$ is an arbitrary
function of $R$, and $\mathcal{L}_{matt}$ is the Lagrangian of
matter fields. Variation with respect to metric $g_{\mu \nu }$,
leads to the field equations
\begin{equation}
{G}_{\mu \nu }=\mathrm{T}_{\mu \nu }^{Curv}+\frac{\mathrm{T}_{\mu \nu
}^{matt}}{F^{\prime }(R)}  \label{FE}
\end{equation}
where ${G}_{\mu \nu }$ is the Einstein tensor and the gravitational
stress-energy tensor is
\begin{equation}
\mathrm{T}_{\mu \nu }^{Curv}=\frac{1}{F^{\prime }(R)}(\frac{1}{2}g_{\mu \nu
}(F(R)-RF^{\prime }(R))+F^{\prime }(R)^{;\alpha \beta }(g_{\alpha \mu
}g_{\beta \nu }-g_{\mu \nu }g_{\alpha \beta }))  \label{Tcurv}
\end{equation}
with $F^{\prime }(R)\equiv dF(R)/dR$ and $\mathrm{T}_{\mu \nu
}^{matt}$ the standard matter stress-energy tensor derived from
the matter Lagrangian $\mathcal{L}_{matt}$ in the action
(\ref{Action}). One can consider geometrical terms in the left
hand side of the field equation, and therefore Eq. (\ref{FE})
reduces to
\begin{equation}
R_{\mu \nu }F^{\prime }(R)-\nabla _{\mu }\nabla _{\nu }F^{\prime }(R)+\left(
\Box F^{\prime }(R)-\frac{1}{2}F(R)\right) g_{\mu \nu }=\mathrm{T}_{\mu \nu
}^{matt}.  \label{FieldEq1}
\end{equation}
The trace of Eq. (\ref{FieldEq1}) reduces to
\begin{equation}
n\Box F^{\prime }(R)+RF^{\prime }(R)-\frac{n+1}{2}F(R)=\mathrm{T}
\label{Trace}
\end{equation}
Here, we consider the special case of $F(R)$ gravity, namely, $R^{N}$
gravity and therefore the field equation (\ref{FieldEq1}) reduces to
\begin{equation}
\left[ NR_{\mu \nu }-\frac{1}{2}g_{\mu \nu }R\right] R^{N-1}+N\left[ g_{\mu
\nu }\Box -\nabla _{\mu }\nabla _{\nu }\right] R^{N-1}=\mathrm{T}_{\mu \nu
}^{matt}  \label{FieldEq2}
\end{equation}
It is easy to show that for $N=1$\bigskip , Eq. (\ref{FieldEq2}) reduces to
familiar Einstein gravity.

\section{Black Hole Solutions of $R^{N}$ gravity without Matter field:\label%
{SolF(R)}}

Here we want to obtain the $(n+1)$-dimensional static solutions of
Eqs. (\ref{FieldEq2}) without any matter field
($\mathcal{L}_{m}=0$, and then $\mathrm{T}_{\mu \nu }^{matt}=0$)
with $N\in
\mathbb{N}
$. We assume that the metric has the following form
\begin{equation}
ds^{2}=-g(r)dt^{2}+\frac{dr^{2}}{g(r)}+r^{2}d\Omega _{k}^{2},  \label{Metric}
\end{equation}
where
\begin{equation}
d\Omega _{k}^{2}=\left\{
\begin{array}{cc}
d\theta _{1}^{2}+\sum\limits_{i=2}^{n-1}\prod\limits_{j=1}^{i-1}\sin
^{2}\theta _{j}d\theta _{i}^{2} & k=1 \\
d\theta _{1}^{2}+\sinh ^{2}\theta _{1}d\theta _{2}^{2}+\sinh ^{2}\theta
_{1}\sum\limits_{i=3}^{n-1}\prod\limits_{j=2}^{i-1}\sin ^{2}\theta
_{j}d\theta _{i}^{2} & k=-1 \\
\sum\limits_{i=1}^{n-1}d\phi _{i}^{2} & k=0
\end{array}
\right. ,  \label{dOmega}
\end{equation}
which represents the line element of an $(n-1)$-dimensional hypersurface
with constant curvature $(n-1)(n-2)k$ and volume $V_{n-1}$. To find the
function $g(r)$, one may use any components of Eq. (\ref{FieldEq2}). The
solution which satisfies all the components of the gravitational field
equations (\ref{FieldEq2}), can be written as
\begin{equation}
g(r)=k-\frac{m}{r^{n-2}}+\frac{\lambda (N-1)}{r^{n-1}}  \label{g(r)}
\end{equation}
where $m$ and $\lambda $ are integration constants which
proportional to the mass and charge parameter respectively. In
order to study the general structure of this solution, we first
look for the curvature singularities. It is easy to show that the
Kretschmann scalar $R_{\mu \nu \alpha \beta }R^{\mu \nu \alpha
\beta }$ diverges at $r=0$, it is finite for $r>0$ and goes to
zero as $r\rightarrow \infty $. Thus, there is an essential
singularity located at $r=0$. The event horizon(s), if there
exists any, is (are) located at the root(s) of $g^{rr}=g(r)=0$
\[
kr_{+}^{n-1}-mr_{+}+\lambda =0.
\]

\begin{figure}[tbp]
\epsfxsize=10cm \centerline{\epsffile{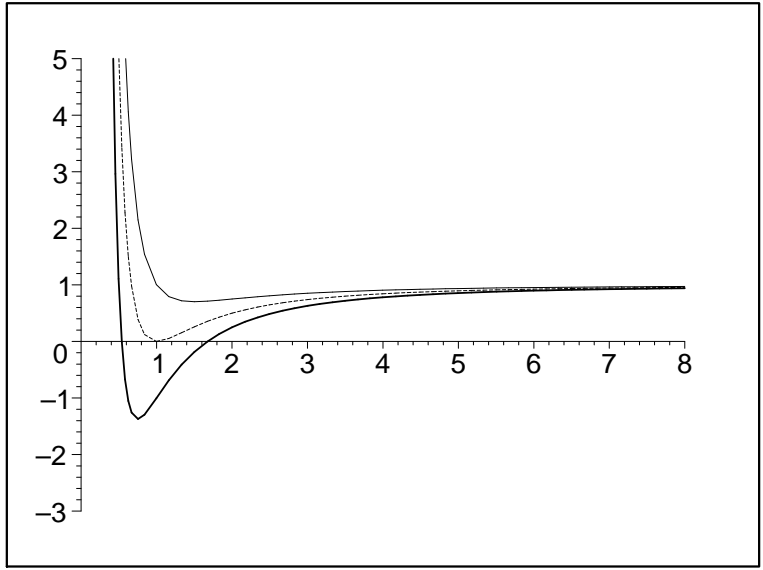}}
\caption{$g(r)$ versus $r$ with for $N=2$, $\protect\lambda =2$, $k=1$, $n=4$
and $m=2<m_{\mathrm{ext}}$ (continuous line), $m=m_{\mathrm{ext}}=3$
(doashed line) and $m=4>m_{\mathrm{ext}}$ (bold line).}
\label{Fig1}
\end{figure}
The temperature may be obtained through the use of the definition of surface
gravity. One obtains
\begin{equation}
T=\frac{g^{\prime }(r_{+})}{4\pi }=\frac{(n-2)mr_{+}-(n-1)\lambda }{4\pi
r_{+}^{n}}=\frac{(n-2)r_{+}^{n-1}-\lambda }{4\pi r_{+}^{n}}  \label{Temp}
\end{equation}
Using the fact that the temperature of the extreme black hole is
zero, it is easy to show that the condition for having an extreme
black hole is that the mass parameter is equal to $m_{ext}$, where
$m_{ext}$ is given as
\begin{equation}
m_{ext}=(n-1)\left( \frac{\lambda }{n-2}\right) ^{(n-2)/(n-1)}.  \label{mext}
\end{equation}
The metric of Eqs. (\ref{Metric}), (\ref{dOmega}) and (\ref{g(r)}) presents
a black hole solution with inner and outer horizons, provided the mass
parameter $m$ is greater than $m_{ext}$, an extreme black hole for $m=m_{ext}
$, and a naked singularity otherwise (see Fig(\ref{Fig1}) for more details).
This behavior is the same as Reissner-Nordstr\"{o}m black holes, but, the
metric function $g(r)$, presented here, differ from the standard
higher-dimensional Reissner-Nordstr\"{o}m solutions since the electric
charge term in the metric function is proportional to $r^{-(n-1)}$ and in
the standard higher dimensional charged black hole solutions is proportional
to $r^{-2(n-2)}$. It is notable that for $N=1$ ($F(R)=R$: Einstein gravity),
the third term in Eq. (\ref{g(r)}) vanishes and the solutions reduce to
Schwarzschild like solutions.

\section{The Solutions of Einstein Gravity in the Presence of Nonlinear
Maxwell Source:\label{ENonMax}}

In this section, we consider the $(n+1)$-dimensional action in which gravity
is coupled to nonlinear electrodynamics field with an action
\begin{equation}
\mathcal{I}_{G}^{\prime }=-\frac{1}{16\pi }\int\limits_{\partial
M}d^{n+1}x\sqrt{-g}\left[ R-\alpha (F_{\alpha \beta }F^{\alpha
\beta })^{s}\right] , \label{IG}
\end{equation}
where $\alpha $ is a coupling constant and the exponent $s$ represented the
nonlinear power of the electromagnetic field. Varying the action (\ref{IG})
with respect to the metric tensor $g_{\mu \nu }$ and the electromagnetic
field $A_{\mu }$, the equations of gravitational and electromagnetic fields
may be obtained as
\begin{equation}
G_{\mu \nu }=\mathrm{T}_{\mu \nu }^{matt},  \label{GravEq}
\end{equation}
\begin{equation}
\partial _{\mu }\left[ \sqrt{-g}F^{\mu \nu }(F_{\alpha \beta }F^{\alpha
\beta })^{s-1}\right] =0,  \label{MaxEq}
\end{equation}
where
\begin{equation}
\mathrm{T}_{\mu \nu }^{matt}=2\alpha \left[ sF_{\mu \rho }F_{\nu }^{\rho
}(F_{\alpha \beta }F^{\alpha \beta })^{s-1}-\frac{1}{4}g_{\mu \nu
}(F_{\alpha \beta }F^{\alpha \beta })^{s}\right]   \label{TNonMax}
\end{equation}
It is easy to show that when $s$ goes to $1$, the Eqs.
(\ref{IG})--(\ref{MaxEq}), reduce to the in Einstein-standard
Maxwell gravity in higher dimensions. The Maxwell equation
(\ref{MaxEq}) with metric (\ref{Metric}) can be integrated
immediately to give
\begin{equation}
F_{tr}=\left\{
\begin{array}{cc}
0, & s=0,\frac{1}{2} \\
-\frac{q}{r}, & s=\frac{n}{2} \\
\frac{(n-2s)q}{(2s-1)r^{(n-1)/(2s-1)}}, & \text{otherwise}
\end{array}
\right. ,  \label{Ftr}
\end{equation}
where $q$, an integration constant where the electric charge of
the spacetime is related to this constant. Inserting the Maxwell
fields (\ref{Ftr}) and the metric (\ref{Metric}) in the field
equation (\ref{GravEq}), one can show that these equations have
the following solutions
\begin{equation}
g(r)=k-{\frac{m}{r^{n-2}}+\alpha \times }\left\{
\begin{array}{cc}
0, & s=0,\frac{1}{2} \\
\frac{(-1)^{3n/2}2^{n/2}q^{n}\ln r}{r^{n-2}}, & s=\frac{n}{2} \\
\frac{(-1)^{s}\left( 2s-1\right) ^{2}}{(n-1)\left( 2s-n\right)
r^{2(ns-3s+1)/(2s-1)}}\left[
\frac{2(2s-n)^{2}q^{2}}{(2s-1)^{2}}\right] ^{s}, &
\text{Otherwise}
\end{array}
\right. ,  \label{g(r)2}
\end{equation}
where $m$ is the integration constant which is related to mass parameter. In
the linear case ($s=1$), the solutions reduce to the higher dimensional
Reissner-Nordstr\"{o}m solutions with linear Maxwell source as they should
be. Straightforward calculation of Kretschmann scalar shows that there is an
curvature singularity located at $r=0$.

Before studying in details the spacetime, we first specify the sign of the
coupling constant $\alpha $ in term of the exponent $s$ in order to ensure a
physical interpretation of our future solutions. In fact, the sign of the
coupling constant in the action (\ref{IG}) can be chosen such that the
energy density, i.e. the $T_{_{\widehat{0}\widehat{0}}}$ component of the
energy-momentum tensor in the ortonormal frame, is positive
\begin{equation}
T_{_{\widehat{0}\widehat{0}}}=(-1)^{s+1}\alpha (2s-1)\left(
2F_{tr}^{2}\right) ^{s}>0.  \label{T00}
\end{equation}
This condition selects two branches depending on the range of the nonlinear
parameter $s$,
\begin{equation}
sgn(\alpha )=\left\{
\begin{array}{cc}
(-1)^{1-s}, & s>\frac{1}{2} \\
(-1)^{-s}, & s<\frac{1}{2}
\end{array}
\right.   \label{sgn}
\end{equation}
while the cases $s=0,1/2$ is excluded because in these cases $F_{tr}$ (and
charge term in (\ref{g(r)2})) vanishes.

Now, we want to investigate the special case, such that the
electromagnetic field equation be invariant under conformal
transformation ($g_{\mu \nu }\longrightarrow \Omega ^{2}g_{\mu \nu
}$ and $A_{\mu }\longrightarrow A_{\mu }$). The idea is to take
advantage of the conformal symmetry to construct the analogues of
the four dimensional Reissner-Nordstr\"{o}m solutions in higher
dimensions. It is easy to show that for Lagrangian in the form
$L(F=F_{\alpha \beta }F^{\alpha \beta })$ in $(n+1)$-dimensions, $
T_{\mu }^{\mu }\propto \left[ F\frac{dL}{dF}-\frac{n+1}{4}L\right]
$; so $ T_{\mu }^{\mu }=0$ implies $L(F)=Constant\times
F^{(n+1)/4}$. Indeed, in our case ($L(F)\propto F^{s}$), for
$s=(n+1)/4$, the Maxwell action (\ref{GravEq} ) enjoys the
conformal symmetry in arbitrary dimensions. In this case the
metric function $f(r)$ and electromagnetic field $F_{tr}$ reduce
to \cite{HendiRastegar}
\begin{equation}
g(r)=k-{\frac{m}{r^{n-2}}+}\frac{2^{(n-3)/4}q^{(n+1)/2}}{r^{n-1}},
\label{gConformal}
\end{equation}
\begin{equation}
F_{tr}=\frac{q}{r^{2}}.  \label{EConformal}
\end{equation}
Here, we calculate the electric charge of the Einstein-conformally invariant
Maxwell solutions. To determine the electric field, we should consider the
projections of the electromagnetic field tensor on special hypersurfaces.
The electric charge per unit volume of the black hole can be found by
calculating the flux of the electromagnetic field at infinity, obtaining
\begin{equation}
Q=\frac{2^{(n-3)/4}(n+1)q^{(n-1)/2}}{16\pi },  \label{Charge}
\end{equation}
which confirm that $q$ is related to the electrical charge of the
spacetime. Comparing Eq. (\ref{g(r)}) with Eq. (\ref{gConformal})
(and let $\lambda =[2^{(n-3)/4}q^{(n+1)/2}]/(N-1)$), show that
there is a correspondence between the Einstein-conformally
invariant Maxwell solutions and the solutions of $F(R)$ gravity
without matter field. One can think about the same effects of \
$\mathrm{T}_{\mu \nu }^{Curv}$ in $F(R)$ gravity (Eq. (\ref {FE})
without $\mathrm{T}_{\mu \nu }^{matt}$) and $\mathrm{T}_{\mu \nu
}^{matt}$ in Einstein-conformally invariant Maxwell gravity
(Eq.(\ref{GravEq})) and therefore $\lambda $ is related to the
charge parameter $q$.

\section{Conclusions}

In the presented paper, we considered the special case of $F(R)$
gravity, so-called $R^{N}$ gravity and obtained its topological
black hole solutions in higher dimensions. We foud that, such as
charged black hole solutions, these solutions may be interpreted
as black hole solutions with two event horizons, extreme black
hole and naked singularity provide that the mass parameter $m$ is
greater than an extremal value $m_{ext}$, $m=m_{ext}$ and
$m<m_{ext}$ respectively. But the presented solutions differ from
the standard higher-dimensional Reissner-Nordstr\"{o}m solutions
since the electric charge term in obtained metric function is
proportional to $r^{-(n-1)}$ and in the standard higher
dimensional charged black hole solutions is proportional to
$r^{-2(n-2)}$.

Next, we presented topological black hole solutions in Einstein
gravity with nonlinear electromagnetic field. Then, we restricted
ourself to the special case, namely, the conformally invariant
Maxwell field coupled to Einstein gravity and discuss about the
correspondence between the Einstein-conformally invariant Maxwell
solutions and the solutions of $F(R)$ gravity without matter field
in arbitrary dimensions. It is easy to show that presented
solutions reduce to Reissner-Nordstr\"{o}m black hole in four
dimension.

\begin{acknowledgements}
It is a pleasure to acknowledge discussions with, and/or comments
by, R. G. Cai, M. H. Dehghani and S. Otarod. This work has been
supported financially by Research Institute for Astronomy and
Astrophysics of Maragha.
\end{acknowledgements}

\end{document}